\documentclass[superscriptaddress,prd,aps,amsmath,amssymb,showpacs,showkeys,11pt]{revtex4-2}
\usepackage[dvips]{graphicx,color}
\usepackage{times}
\usepackage{tensor}
\usepackage[symbol]{footmisc}
\usepackage{subfigure}

\usepackage{xcolor}
\usepackage[%
  colorlinks=true,
  urlcolor=blue,
  linkcolor=red,
  citecolor=blue
]{hyperref}
\usepackage{orcidlink}
\DeclareUnicodeCharacter{2212}{-}

\begin{document}

\color{black}       

\title{Dark matter effects of a black hole with nonsingular Yukawa--modified potential in Einstein‐-Gauss‐-Bonnet Gravity}

\author{Yassine Sekhmani\orcidlink{0000-0001-7448-4579}}
\email[Email: ]{sekhmaniyassine@gmail.com}
\affiliation{Ratbay Myrzakulov Eurasian International Centre for Theoretical Physics, Astana 010009, Kazakhstan.}
\affiliation{L. N. Gumilyov Eurasian National University, Astana 010008,
Kazakhstan.}
\author{A. A. Araújo Filho\orcidlink{0000-0002-8790-3944}}
\email[Email: ]{dilto@fisica.ufc.br}
\affiliation{Departamento de Física, Universidade Federal da Paraíba, Caixa Postal 5008, 58051-970, João Pessoa, Paraíba,  Brazil}

\author{Ratbay Myrzakulov \orcidlink{0000-0002-5274-0815}}
\email[Email: ]{rmyrzakulov@gmail.com }
\affiliation{L. N. Gumilyov Eurasian National University, Astana 010008,
Kazakhstan.}
\affiliation{Ratbay Myrzakulov Eurasian International Centre for Theoretical
Physics, Astana 010009, Kazakhstan.}

\author{Adam Z. Kaczmarek \orcidlink{0000-0003-0956-8623}}
\email[Email: ]{adamzenonkaczmarek@gmail.com }

\affiliation{Institute of Physics, Faculty of Science and Technology, Jan D{\l}ugosz University in Cz{\c{e}}stochowa, 13/15 Armii Krajowej Ave., 42200 Cz{\c{e}}stochowa, Poland}
\author{ Javlon Rayimbaev\orcidlink{0000-0001-9293-1838}}
\email[Email: ]{javlon@astrin.uz}
\affiliation{
New Uzbekistan University,  Movarounnahr Str. 1, Tashkent 100007, Uzbekistan}
\affiliation{
Central Asian University, Tashkent 111221, Uzbekistan}
\affiliation{
University of Tashkent for Applied Sciences, Gavhar Str. 1, Tashkent 100149, Uzbekistan }
\affiliation{Institute of Fundamental and Applied Research, National Research University TIIAME, Kori Niyoziy 39, Tashkent 100000, Uzbekistan}
\author{Dominik Szcz{\c{e}}{\'s}niak\orcidlink{0000-0003-1880-1255}}
\email[Email: ]{d.szczesniak@ujd.edu.pl}
\affiliation{Institute of Physics, Faculty of Science and Technology, Jan D{\l}ugosz University in Cz{\c{e}}stochowa, 13/15 Armii Krajowej Ave., 42200 Cz{\c{e}}stochowa, Poland}

\begin{abstract}
This paper investigates the contribution of the nonsingular Yukawa--modified potential in the context of four--dimensional Einstein--Gauss--Bonnet (EGB) gravity modeling by a static and spherically symmetric black hole solution. These Yukawa--type corrections are essentially described along two parameters, $\beta$ and $\lambda$, affecting Newton's law of gravity at large distances, and a deformation parameter $\ell_0$, which is essential at short distances. Primarily, the strongest effect is encoded in $\beta$, which alters the total mass of the black hole with additional mass proportional to $\beta\mathcal{M}$, imitating the effects of dark matter at large distances from the black hole. In contrast, the effect due to $\lambda$ is small for astrophysical values. On the other hand, the EGB gravity is ruled by the Gauss--Bonnet (GB) coupling constant $\alpha$, a fundamental parameter of the theory. We pay particular attention to thermodynamic stability, critical orbits, geodesics and quasinormal modes. The results demonstrate stability of the black hole solution for a range of values of the GB coupling constant $\alpha$. Furthermore, this study investigates the null geodesic motion, namely the shadow behavior, providing intriguing results in relation to the size of the black hole shadow.

\end{abstract}

\maketitle

\section{Introduction}
The idea to modify and provide alternatives to general relativity (GR) is as old as the original formulation of the theory. The first of such approaches dates back to the times of Einstein and was presented by Weyl, Eddington, Kaluza and Klein \cite{Weyl:1918pdp,Kaluza:1921tu,eddington1924,Klein:1926fj}. These extensions are relevant even now, since the standard formulation of the GR faces serious challenges when considering the dark sector (the dark energy and the dark matter) \cite{Clifton:2011jh}. In this respect, one of the extended theories of particular convenience is the Einstein-Gauss-Bonnet (EGB) gravity, historically introduced by Lovelock \cite{Lovelock:1971yv,Fernandes:2022zrq}. The pivotal aspect of this theory is related to the Gauss-Bonnet (GB) term
\begin{equation}  
\mathcal{G}=R^2-4R^{\mu\nu}R_{\mu\nu}+R^{\mu\nu\rho\sigma}R_{\mu\nu\rho\sigma},\label{E1}\end{equation}
which, although associated with the higher-curvature corrections in the Einstein-Hilbert action, leads to field equations that are at most of the second order in metric. Thus, Ostrogardski's instabilities are avoided despite the introduction of the GB term \cite{Fernandes:2022zrq}. However, while important in the five-dimensions (or in more general couplings such as the $f(\mathcal{G})$ models), the GB term in its linear form ($D\rightarrow 4$ dimensions) vanishes at the level of the field equations and becomes {\it de facto} irrelevant \cite{Bonifacio:2020vbk,Fernandes:2022zrq}. Still, by using the proper rescaling ($\alpha \rightarrow \frac{\alpha}{D-4}$), the resulting equations in four dimensions may have the GB contribution, in contrast to Lovelock's theorem \cite{Glavan:2019inb}. As a consequence, recent years have witnessed a resurgence of Gauss-Bonnet gravity in its four-dimensional form. Among many, one of the most interesting and insightful studies was provided in the context of black holes \cite{Yang:2020jno,Konoplya:2020bxa,Konoplya:2020qqh,Ghosh:2020syx,Wei:2020poh,Zhang:2020qam,Yerra:2022eov,Singh:2022dth,Biswas:2022qyl,Errehymy:2023xpc,Carvalho:2022ywl}. More precisely, the latest activities in the context of EGB gravity have given rise to a wealth of investigations into thermodynamics \cite{Belhaj:2022qmn}, quasinormal modes (QNMs) \cite{Gogoi:2023ffh, Sekhmani:2023ict}, or shadows behavior \cite{Belhaj:2022ntd, Belhaj:2022kek, Belhaj:2022gcj, Belhaj:2023nhq, Gogoi:2023ntt}. These investigations lend particular interest to the EGB gravity; in particular, this theory is linked to the effective theories of superstrings, which are encoded via the GB coupling constant.

In general, the notion of a black hole (BH) plays a key role in the development of proper spacetime description, establishing one of the core research directions in modern physics \cite{Bronnikov:2012wsj,Moffat:2014aja,Moffat:2015kva,Dehghani:2017szv,Murk:2021qiv}. This concept confronts the existing paradigms in both general relativity and the quantum realm via enigmatic features like the singularities or the information paradox, highlighting the limits of classical relativity and the necessity for a quantum gravity framework \cite{Chen:2014xgj,Raju:2020smc}. It also constitutes an attractive tool to study the aforementioned dark sector \cite{Filho:2023abd,jusufi2024charged}, which is notoriously hard to detect due to the lack of electromagnetic interaction. This fact appears to be especially intriguing in the context of the recently introduced models based on the Yukawa force \cite{Berezhiani:2009kv,Borka:2013dba,Garny:2015sjg,Arvanitaki:2016xds,Jusufi:2023xoa}. Approaches of that kind are closely related to the modified Newtonian dynamics \cite{Milgrom:1983md}, as the potential of the Yukawa type alters the Newtonian one in a similar manner \cite{Jusufi:2023xoa}. Recent breakthroughs point to dark matter being explained by Yukawa coupling between baryonic matter in terms of long-range interaction \cite{Berezhiani:2009kv,Jusufi:2023xoa}. Nevertheless, it should be noted that these ideas are not yet properly addressed by modified theories of gravity (such as EGB gravity) \cite{Chabab:2020xwr,Maier:2021jxv,Filho:2023abd}.

Motivated by the above, we attempt to investigate the relationship between the BH region and the specific dark matter model within the framework of the four-dimensional (4D) EGB gravity \cite{Fernandes:2022zrq}. This is done by introducing a new BH solution supported by the Yukawa corrections. As a result, we are able to provide a novel contribution to the fundamental understanding of the BH spacetime in the dark sector environment, supplementing and extending previous related studies \cite{Saurabh:2020zqg,Yerra:2022eov,Errehymy:2023xpc,Filho:2023abd}. These results are followed by a discussion of the possible related effects. In particular, our study focuses on investigating shadow and light ray behavior in Yukawa-corrected BH spacetime by using 4D EGB gravity. The BH shadows are unique images produced by the light deflected near the event horizon, creating the perfect opportunity to study the connection between various spacetimes and geometries. This includes analysis of the dark sector and the dark matter candidates \cite{Cunha:2015yba,Klimenko:2017vfq,Hou:2018avu,Haroon:2018ryd,Jusufi:2020cpn,Filho:2023abd,Arbey:2021gdg,Anjum:2023axh} as well as the discussion of the QNMs closely linked to the radius of the BH shadows \cite{Kokkotas:1999bd,Moss:2001ga,Jusufi:2020dhz,Konoplya:2020bxa,Atamurotov:2022nim}. Hence, through geometric optics, our work explores the intricate relationship between the BH geometry, QNMs, and the modified Yukawa potential following recent trends \cite{Liu:2020ola,Ghasemi-Nodehi:2020oiz,Jusufi:2020odz,Cai:2021ele,Jafarzade:2020ova,Saurabh:2020zqg,Campos:2021sff,Atamurotov:2022iwj,Chen:2022nbb,Rayimbaev:2022mrk,Dymnikova:2023opi,Turimov:2023xxm,Vishvakarma:2023csw,Ghosh:2022gka,Gonzalez:2023rsd}. We argue that such investigations are essential for understanding the potential influence of dark matter on future BH shadow observations \cite{Perlick:2021aok,Akiyama:2022xqj,Atamurotov:2022iwj}.

The presented work is organized as follows: (i) The theoretical background and description of the Yukawa corrections to the gravitational potential are introduced in Section II. Therein, the new BH solution for the 4D EGB with the Yukawa potential is obtained, and the physical characteristics of such a model are thoroughly discussed \cite{Berezhiani:2009kv}. (ii) In the Section IV of the manuscript, the critical orbits and geodesics are analyzed. (iii) This is followed by investigations of the BH shadows and related properties in Section V. (iv) To this end, in Section VI, quasinormal modes are thoroughly discussed \cite{Matyjasek:2017psv,Cunha:2019hzj,Filho:2023abd}. (vi) The last part is devoted to the conclusions and summary of the presented work, including indications of new perspectives that this manuscript provides.

\section{BLACK HOLE SOLUTION WITH YUKAWA POTENTIAL}
Motivated by the exciting black hole solution with the dark matter effect recently studied in A, we are looking forward to extending the situation into the EGB gravity framework.

We consider the EGB action with a negative cosmological constant as
  \begin{equation}
S=\frac{1}{16\pi}\int \mathrm{d}^Dx\sqrt{-g}\left(R-2\Lambda+\mathcal{G}\right),
\label{E2}
\end{equation} 
where $R$ is the Ricci scalar, $\mathcal{G}$ is mentioned in Eq. (\ref{E1}) and $\alpha$ is the GB coupling constant. In the rest of this work, we assume the $4D$ static and spherically symmetric spacetime anstaz given by
\begin{equation}
\mathrm{d} s^2=-f(r) \mathrm{d}t^2+\frac{\mathrm{d}r^2}{f(r)}+r^2(\mathrm{d}\theta^2+ \sin^2\theta\mathrm{d}\phi^2).
\label{E3}
\end{equation}
Even though in the $4D$ scenario, the GB term is practically a total derivative and therefore does not contribute to gravitational dynamics, an additional scalar field can be coupled to the GB term, which is referred to as Einstein--Dilaton Gauss--Bonnet theory influencing that dynamic \cite{Kanti:1997br,Kanti:1995vq}. To overcome the topological problem, Glavan and Lin \cite{Glavan:2019inb} investigated the $4D$ set--in of the GB term, considering a rescaling of the coupling constant 
\begin{equation}
    \alpha \rightarrow \frac{\alpha}{D-4}\label{E4}.
\end{equation}
Consequently, dealing with the $4D$ case is now achieved, and the field equation for component 00, incorporating a material source, results
 \begin{align}
    \frac{1}{r}\frac{\mathrm{d}f}{\mathrm{d}r}+\frac{f}{r^2}-\frac{1}{r^2}&-\alpha\bigg(\frac{2(f-1)}{r^3}\frac{\mathrm{d}f}{\mathrm{d}r}-\frac{(f-1)^2}{r^4}\bigg)+\Lambda= -\rho(r),\label{E5}
\end{align}
where $T_0^0 =-\rho(r)$.

In what follows, an examination of the matter source regarding the modified Yukawa potential is required. This contribution was recently discovered to probe the cosmological significance of the modified part of the Yukawa potential \cite{Gonzalez:2023rsd}. For that reason, the gravitational potential we considered is modified by the regular Yukawa–type potential given as follows
\begin{equation}
    \Phi(r) =\frac{-G \mathcal{M}\, m}{\sqrt{r^{D-2}+ \ell_0^{D-2}}}\left(1 +\beta\,e^{-\frac{r}{\lambda}}\right).\label{EE5}
\end{equation}
Notably, the wavelength of massive graviton is found to be $\lambda =\frac{\hbar}{m_g c} $, and $\ell_o$ represents a deformed parameter of Planck length size. It is well known that the energy density of the modified matter can be explicitly derived by considering $\rho(r) =\frac{1}{4\pi}\Delta \Phi(r)$. Taking this element into account, we obtain the following energy density:
\begin{equation}
 \rho(r) =  \frac{e^{-\frac{r}{\lambda}}\beta\, \mathcal{M}}{4\pi\,r\lambda^2(r^2+ \ell_0^2)^{5/2}}\mathcal{A}+\frac{3\mathcal{M}\ell_0^2}{4\pi(r^2 +\ell_0^2)^{5/2}}\label{E7}
\end{equation}
where on has, $\mathcal{A}=2\lambda\ell_0^4+(3\lambda^2\ell_0^2-\ell_0^4)r+2\lambda\ell_0^2 r^2-2\ell_0^2r^3-r^5$. A closer interpretation of the energy density expression reveals that the two terms have different characteristics: the first term is linearly proportional to $\beta$ and provides interesting results at large distances, while the second is proportional to $\ell_0$ and plays particular attention to short distances. Specifically, if we ignore long--range modification, that is, as a particular case of our results, only the second term maintains consistency with \cite{Nicolini:2019irw}.

To proceed with finding a BH solution within EGB gravity, one can expand the first term of the energy density (\ref{E7}) in a series around $\ell_0$. This can result in the following expression:
    \begin{equation}
 \rho(r) =  -\frac{\beta\, \mathcal{M}}{4\pi\,r\lambda^2}e^{-\frac{r}{\lambda}}+\frac{3\mathcal{M}\ell_0^2}{4\pi(r^2 +\ell_0^2)^{5/2}}+ \mathcal{O}(\ell_0^2\,\beta)
\end{equation}
Accordingly, it may be possible to distinguish the physical nature of the current energy density. Indeed, the first term corresponds to that of \cite{Gonzalez:2023rsd}. In addition, it is worth noting that the negative sign indicates that the energy conditions are violated inside the black hole. Alternatively, we further conjecture that the field equations with a cosmological constant remain valid, i.e., $G_{\mu\nu} + \Lambda g_{\mu\nu} + \alpha H_{\mu\nu} = 8\pi T_{\mu\nu}$, whereby the effects of effective dark matter are encoded in the stress-energy tensor. And, from the gravitational field equations, one can obtain
    \begin{align}
        \frac{1}{r}\frac{\mathrm{d}f}{\mathrm{d}r}&+\frac{f}{r^2}-\frac{1}{r^2}-\alpha\bigg(\frac{2(f-1)}{r^3}\frac{\mathrm{d}f}{\mathrm{d}r}-\frac{(f-1)^2}{r^4}\bigg)+\Lambda -\frac{2\beta\, \mathcal{M}}{r\lambda^2}e^{-\frac{r}{\lambda}}+\frac{6\mathcal{M}\ell_0^2}{(r^2 +\ell_0^2)^{5/2}}=0
    \end{align}
This equation yields a pair of distinct solutions denoted by the signs $\pm$. These solutions could be exposed as follows:
    \begin{equation}
    f(r)_{+,-}=1+\frac{r^2}{2\alpha}\Biggr\lbrace{1\pm\sqrt{1+\frac{4}{3}\alpha\bigg(6\mathcal{M}\bigg(\frac{1}{(\ell_0^2+r^2)^{3/2}}+\frac{e^{-\frac{r}{\lambda}}\beta(r+\lambda)}{r^3\lambda}\bigg)+\Lambda+\frac{3\alpha}{r^3}c_1\bigg)}\Biggr\rbrace},
\end{equation}
where $c_1$ is an integration constant. Concretely, the physical solution is that with a negative branch, which brings to the ordinary BH solutions in the context of GR considering the limit $\alpha\rightarrow0$. Practically speaking, the negative branch is therefore a new space-time generalization in the sense of modifying the gravity of the study in Ref. \cite{Gonzalez:2023rsd}. Moreover, we can interpret the physical content of the obtained BH solution in such a way that the second term carries each modification of the geometry due to $\ell_0$. Further, the third term consists entirely of the apparent effect of dark matter, while the fourth term is due to the contribution of the cosmological constant \cite{Nicolini:2019irw}. It is worth noting that the vanishing of the third, fourth, and last terms, the BH solution in this case, can be brought to that of Ref. \cite{Gonzalez:2023rsd} with the consideration of the limit $\alpha\rightarrow 0$. Notably, $\ell_0$ is of Planck length order, i.e., $\ell_0 \sim10^{−35}m$ \cite{Nicolini:2019irw}. In large distances and astrophysical black holes with a large mass $\mathcal{M}$, we should set $r \gg\ell_0$; then, the ignorance of $\ell_0$ is constrained and based on this bound conjecture. Setting $c_1=0$ can provide the exact solution of the BH as follows:   
\begin{equation}
    f(r)=1+\frac{r^2}{2\alpha}\Biggr\lbrace{1-\sqrt{1+\frac{4}{3}\alpha\bigg(6\mathcal{M}\bigg(\frac{1}{(\ell_0^2+r^2)^{3/2}}+\frac{e^{-\frac{r}{\lambda}}\beta(r+\lambda)}{r^3\lambda}\bigg)+\Lambda\bigg)}\Biggr\rbrace},
\end{equation}
which generalizes the BH solution in the frame of the EGB gravity model. It is useful to address the relevant physical aspect by working with the physical mass, $M= \mathcal{M}(1 +\beta)$, and this could be presented in the exact solution as follows 
    \begin{equation}\label{solution}
    f(r)=1+\frac{r^2}{2\alpha}\Biggr\lbrace{1-\sqrt{1+\frac{4}{3}\alpha\bigg(\frac{6M}{1+\beta}\bigg(\frac{1}{(\ell_0^2+r^2)^{3/2}}+\frac{e^{-\frac{r}{\lambda}}\beta(r+\lambda)}{r^3\lambda}\bigg)+\Lambda\bigg)}\Biggr\rbrace},
\end{equation}\label{metric}
\begin{figure*}[tbh!]
      	\centering{
       \includegraphics[scale=0.73]{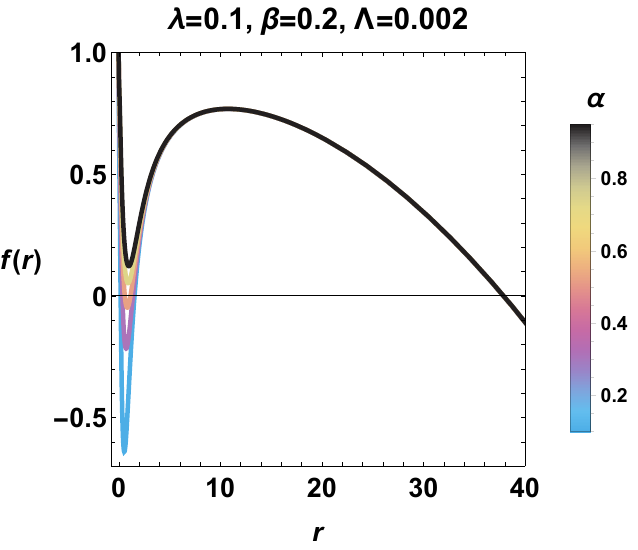} \hspace{2mm}
      	\includegraphics[scale=0.73]{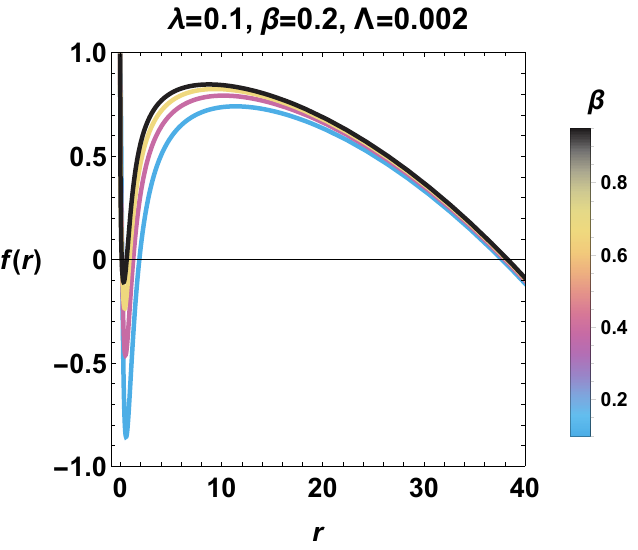} \hspace{2mm}
      }
       
      	\caption{Variation of the black hole metric function (\ref{solution}) with respect to $r$ for various values of the parameter space and with $M=1$ and $\ell_0=0.2$.}
      	\label{fig1}
      \end{figure*}
which, in the context of GR, reduces to another defined exact solution \cite{Gonzalez:2023rsd}.
\begin{equation}
    f(r)=1-\frac{2M r^2}{(1 +\beta)(\ell_0^2+r^2)^{3/2}}-\frac{2M \beta(r+\lambda)e^{-\frac{r}{\lambda}}}{r\lambda(1 +\beta)}-\frac{\Lambda r^2}{3}.
\end{equation}

 To gain a better understanding of the black hole metric (\ref{solution}), Fig. $\ref{fig1}$ offers a suitable graphical analysis. The observation is mostly significant for the parameters $\beta$ and $\alpha$. In the literature on black hole physics, the possible roots of the metric function are classified into two kinds: the smallest root is related to two black hole horizons, and the largest root is in relation to the cosmological horizon. It is observed that such a variation of the parameters $\alpha$ and $\beta$ generates various configurations of the inner horizon, the event and cosmological being closely located at the same horizon radius. A closer observation shows that depending on both the parameters $\alpha$ and $\beta$, the possible set of the horizon radius can reduce from three possible horizon radius up to one, namely the cosmological horizon.

To conduct some analysis of nature's singularities in relation to the physical solution $(\ref{solution})$, the Ricci scalars ($R$), the Ricci square ($R_{\mu\nu}R^{\mu\nu}$) and Kretshmann scalars ($R_{\mu\nu\lambda\sigma}R^{\mu\nu\lambda\sigma}$) are needed
  \begin{figure}[htb!]
    \centering
    \begin{subfigure}
        \centering
        \includegraphics[scale=0.599]{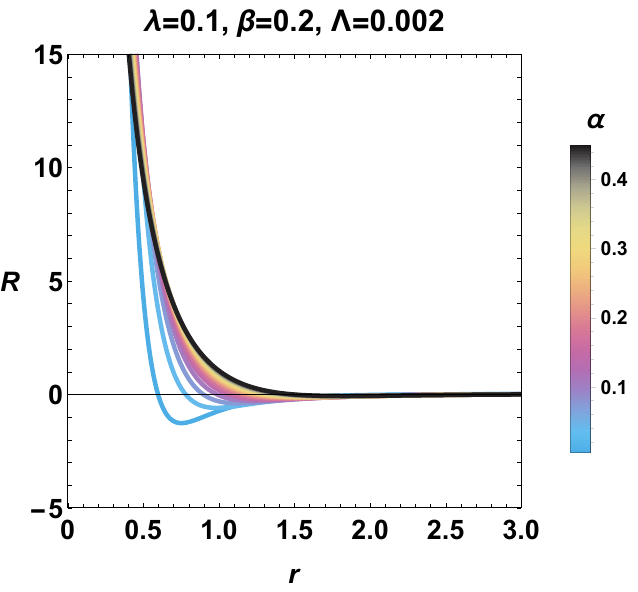}
    \end{subfigure}%
    \hfill
    \begin{subfigure}
        \centering
        \includegraphics[scale=0.68]{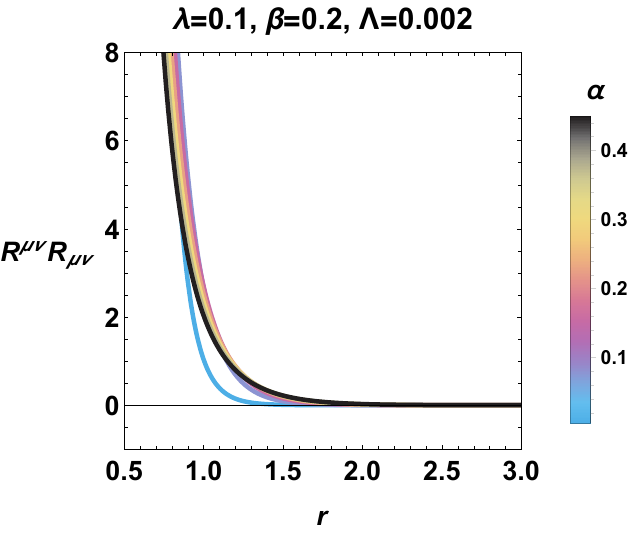}
    \end{subfigure}%
    \\
    \begin{subfigure}
        \centering
        \includegraphics[scale=0.75]{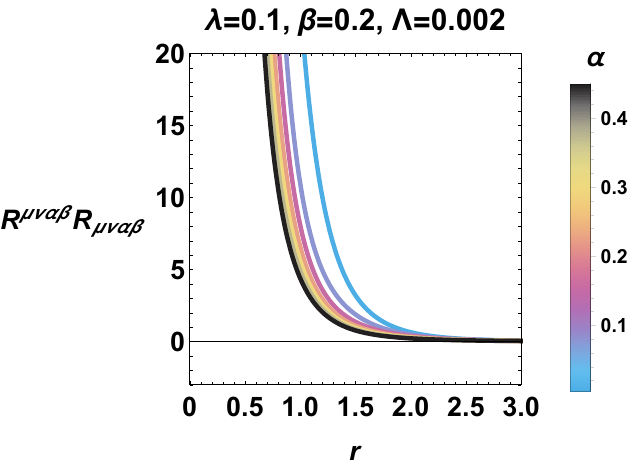}
    \end{subfigure}
    \caption{Variation of Ricci, Ricci squared, and Kretschmann scalar w.r.t. r for various values of $\alpha$ and with $M=1$ and $\ell_0=0.2$.}
    \label{fig2}
\end{figure}
\begin{widetext}
\begin{align}\label{r27}
R&=-\frac{r^2 f''(r)+4 r f'(r)+2 f(r)-2}{r^2},\\
R_{\mu\nu}R^{\mu\nu}&=\frac{r^4 f''(r)^2+8 r^2 f'(r)^2+8 f(r) \left(r f'(r)-1\right)+4 r f'(r) \left(r^2 f''(r)-2\right)+4 f(r)^2+4}{2 r^4},\\
R_{\mu\nu\alpha\beta}R^{\mu\nu\alpha\beta}&=f''(r)^2+\frac{4 f'(r)^2}{r^2}+\frac{4(f(r)-1)^2}{r^4}.
\end{align}
\end{widetext}
Using the metric function in our case, one can see that the scalars are not free from singularity issues. This is mainly due to the presence of the violation of stress--energy conservation.

To offer a thorough discussion of the characteristics of the black hole solution from the standpoint of scalar invariants, Fig. $\ref{fig2}$ provides the relevant features. It is observed that all scalar invariants have similar behaviors and are positive-definite, while negative at a small $r$ for certain valued parameters. The presentation (Fig. $\ref{fig2}$) proves that the Gauss--Bonnet coupling $\alpha$ affects the variation of the Ricci scalar, the Ricci scalar squared, and the Kretshmann. In addition, it is well noted that the three scalars all go to zero when $r$ lies outside the event horizons. Thus, the black hole solution is therefore completely unique in nature and exhibits a physical singularity at $r = 0$, beyond avoidance.


\section{Thermodynamics stability}
Revealing the local thermal stability of the BH solution is the main focus of this section. Proceeding with the thermodynamics aspect is first examined by approaching the solution set at the horizon radius of the equation $f(r_h)=0$, giving an appropriate expression of the black hole mass as
\begin{equation}\label{mass}
    \mathcal{M}=-\frac{(\beta +1) \left(\Lambda  r_h^4-3 \left(\alpha +r_h^2\right)\right)}{6 \left(\frac{r_h^4}{\left(r_h^2+\ell_0^2\right){}^{3/2}}+\frac{\beta  r_h \left(r_h+\lambda
   \right)}{\lambda }e^{-\frac{r_h}{\lambda }} \right)}
\end{equation}
On the other hand, the corresponding Hawking temperature for the metric given by Eq. (\ref{solution}) is \cite{Wald:1984rg}
\begin{equation}
    T_H=\frac{1}{4\pi}\left(\frac{\mathrm{d}f}{\mathrm{d}r}\right)_{r=r_h}
\end{equation}
or, in terms of the parameter space of the black hole system, this quantity is expressed as follows:
\begin{equation}\label{tem}
    T_H=-\frac{1}{4 \pi }\Biggl\{\frac{\left(\Lambda  r_h^4-3 \left(\alpha +r_h^2\right)\right) \bigg(\frac{3
   r_h^5}{\left(\ell_0^2+r_h^2\right)^{5/2}}+\frac{\beta  e^{-\frac{r_h}{\lambda }} \left(3
   \lambda ^2+r_h^2+3 \lambda  r_h\right)}{\lambda ^2}\bigg)}{3 \left(2 \alpha
   +r_h^2\right) \bigg(\frac{r_h^4}{\left(\ell_0^2+r_h^2\right)^{3/2}}+\frac{\beta  r_h
   e^{-\frac{r_h}{\lambda }} (\lambda +r_h)}{\lambda }\bigg)}+\frac{2}{r_h}\Biggr\}
\end{equation}
Accordingly, specific limits on the parameter space can reduce the corresponding Hawking temperature to other defining ones; particularly, the choice leads to the Hawking temperature of the black hole.

Providing the entropy expression is another task for examining the fully thermodynamic framework, for which the first law of black hole thermodynamics is to be applied. Thus, the application law clearly defines the desired expression, namely $\mathrm{d}S =(1-\phi_M)\mathrm{d}M/T_H$. So, using Eq. (\ref{mass}) and Eq. (\ref{tem}) leads to 
\begin{figure*}[tbh!]
      	\centering{
       \includegraphics[scale=0.70]{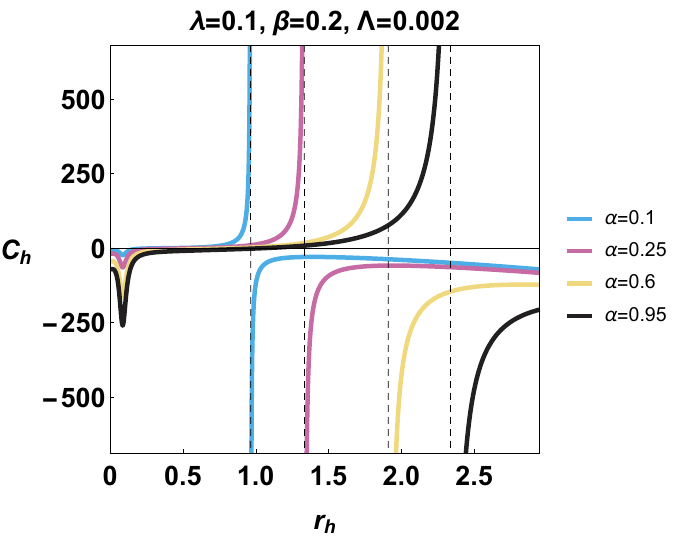} \hspace{2mm}
      	\includegraphics[scale=0.70]{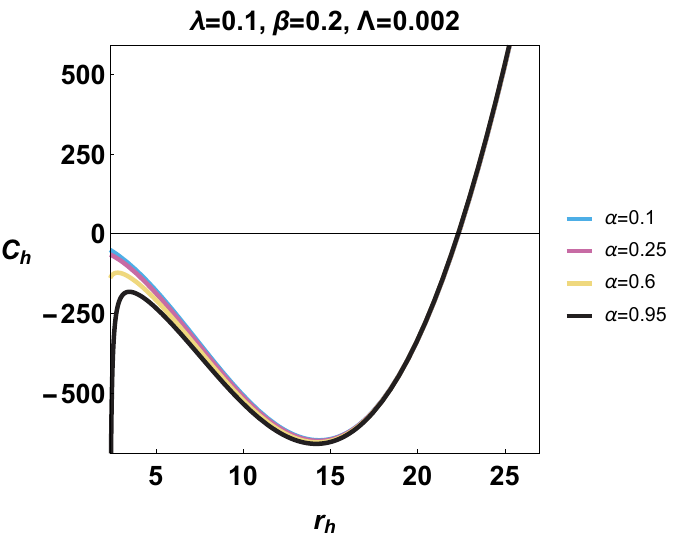} \hspace{2mm}
      }
       
      	\caption{Variation of the Heat capacity $C_h$ (\ref{heat}) with respect to $r_h$ for various values of $\alpha$. }
      	\label{fig3}
      \end{figure*}
\begin{equation}
    S=\pi  \left(r_h^2+4 \alpha  \log (r_h)\right)
\end{equation}
where a closer observation shows that the obtained entropy does not obey the area law, and the corrected factor is given explicitly as
\begin{equation}
    \phi_M=4\pi\int r^2\frac{\partial T_0^0}{\partial M}=1- \frac{r^3}{(\beta +1) \left(\ell_0^2+r^2\right)^{3/2}}-\frac{\beta  e^{-\frac{r}{\lambda
   }} (\lambda +r)}{(\beta +1) \lambda }.
\end{equation}
For further details on the arguments concerning the logarithmic behavior of entropy, refer to \cite{Cai:2009ua}.

In order to predict the local thermal black hole stability, the heat capacity shows, based on the analysis of its sign, whether the black hole system is locally stable or unstable \cite{c1,c2,c3,c4,c5}. We can calculate the heat capacity using the expression $C_h=\mathrm{d}M/\mathrm{d}T_H$. In concrete terms, focusing on the heat capacity demands and employing Eqs. (\ref{mass}) and (\ref{tem}), we obtain the following:
\begin{equation}\label{heat}
    C_h=-2 \pi  (\beta +1)\frac{C_1}{C_2}
\end{equation}
where
\begin{align}
 C_1 &= \Biggl\{\left(4 \Lambda  r_h^3-6 r_h\right)
   \left(-\frac{r_h^4}{\left(\ell_0^2+r_h^2\right)^{3/2}}-\frac{\beta  r_h
   e^{-\frac{r_h}{\lambda }} (\lambda +r_h)}{\lambda }\right)+\left(\Lambda  r_h^4-3
   \left(\alpha +r_h^2\right)\right)
   \bigg(\frac{r_h^5}{\left(\ell_0^2+r_h^2\right)^{5/2}}\nonumber\\
   &+\frac{4 \ell_0^2
   r_h^3}{\left(\ell_0^2+r_h^2\right)^{5/2}}+\frac{\beta  e^{-\frac{r_h}{\lambda }}
   \left(\lambda ^2-r_h^2+\lambda  r_h\right)}{\lambda
   ^2}\bigg)\Biggr\}\nonumber
   \end{align}
   \begin{align}
    C_2&=\left(\frac{r_h^4}{\left(\ell_0^2+r_h^2\right)^{3/2}}+\frac{\beta  r_h
   e^{-\frac{r_h}{\lambda }} (\lambda +r_h)}{\lambda }\right)^2 \Biggl\{\frac{r_h \left(\Lambda 
   r_h^4-3 \left(\alpha +r_h^2\right)\right) \left(\frac{15 \ell_0^2
   r_h^3}{\left(\ell_0^2+r_h^2\right)^{7/2}}-\frac{\beta  e^{-\frac{r_h}{\lambda }} (\lambda
   +r_h)}{\lambda ^3}\right)}{\left(2 \alpha +r_h^2\right)
   \left(\frac{r_h^4}{\left(\ell_0^2+r_h^2\right)^{3/2}}+\frac{\beta  r_h e^{-\frac{r_h}{\lambda
   }} (\lambda +r_h)}{\lambda }\right)}\nonumber\\
   &-\frac{2 r_h
   \left(\Lambda  r_h^4-3 \left(\alpha +r_h^2\right)\right) \left(\frac{3
   r_h^5}{\left(\ell_0^2+r_h^2\right)^{5/2}}+\frac{\beta  e^{-\frac{r_h}{\lambda }} \left(3
   \lambda ^2+r_h^2+3 \lambda  r_h\right)}{\lambda ^2}\right)}{\left(2 \alpha
   +r_h^2\right)^2 \left(\frac{r_h^4}{\left(\ell_0^2+r_h^2\right)^{3/2}}+\frac{\beta  r_h
   e^{-\frac{r_h}{\lambda }} (\lambda +r_h)}{\lambda }\right)}-\frac{6}{r_h^2}\nonumber\\
   &+\frac{\left(4 \Lambda  r_h^3-6 r_h\right)
   \left(\frac{3 r_h^5}{\left(\ell_0^2+r_h^2\right)^{5/2}}+\frac{\beta  e^{-\frac{r_h}{\lambda
   }} \left(3 \lambda ^2+r_h^2+3 \lambda  r_h\right)}{\lambda ^2}\right)}{\left(2
   \alpha +r_h^2\right) \left(\frac{r_h^4}{\left(\ell_0^2+r_h^2\right)^{3/2}}+\frac{\beta  r_h
   e^{-\frac{r_h}{\lambda }} (\lambda +r_h)}{\lambda }\right)}-\frac{\left(\Lambda 
   r_h^4-3 \left(\alpha +r_h^2\right)\right)
  }{\left(2 \alpha
   +r_h^2\right) \left(\frac{r_h^4}{\left(\ell_0^2+r_h^2\right)^{3/2}}+\frac{\beta  r_h
   e^{-\frac{r_h}{\lambda }} (\lambda +r_h)}{\lambda }\right)^2}\nonumber\\
   &\times \left(\frac{r_h^5}{\left(\ell_0^2+r_h^2\right)^{5/2}}+\frac{4 \ell_0^2
   r_h^3}{\left(\ell_0^2+r_h^2\right)^{5/2}}+\frac{\beta  e^{-\frac{r_h}{\lambda }}
   \left(\lambda ^2-r_h^2+\lambda  r_h\right)}{\lambda ^2}\right) \nonumber\\
   &\times\left(\frac{3r_h^5}{\left(\ell_0^2+r_h^2\right)^{5/2}}+\frac{\beta  e^{-\frac{r_h}{\lambda }} \left(3
   \lambda ^2+r_h^2+3 \lambda  r_h\right)}{\lambda ^2}\right)\Biggr\}\nonumber
\end{align}

To gain a better understanding of the heat capacity behavior and results in the local state thermal stability, Fig. \ref{fig3} provides a suitable analysis. It is observed that the system presents two physical limitation points, namely root of the heat capacity $(C_h=0)$, On the other hand, the system exhibits one divergent point for all the valued parameter spaces. Overall, these points involve four regions with specific signs of heat capacity, either negative, which indicates the black hole system is locally thermally stable, or positive, indicating a locally unstable black hole. Roughly speaking, among all the first-- or second--order phase transitions, the system remains locally thermally stable.


\section{Critical orbits and geodesics}

Achieving a thorough comprehension of particle dynamics and the intricate behavior of light rays in the proximity of black hole structures demands a profound understanding of critical orbits. These orbits play a pivotal role in elucidating the unique properties of spacetime, especially when influenced by the effects of dark matter in our specific context.

In our quest to better grasp the impact of the photon sphere, commonly known as the critical orbit, within our black hole scenario, we will leverage the Lagrangian method for computing null geodesics. This method promises a more comprehensive and transparent perspective than the previously introduced geodesic equation. Our analysis seeks to uncover how the mass of the black hole influences the photon sphere, thereby shedding light on the gravitational effects embedded in the black hole solution with Yukawa potential in the context of Gauss--Bonnet. Thereby, we write:
\begin{equation}
\mathcal{L} = \frac{1}{2} g_{\mu\nu}\Dot{x}^{\mu}\Dot{x}^{\nu}.
\end{equation}
When a fixed angle of $\theta=\pi/2$ is taken into account, the previously mentioned expression undergoes simplification, yielding:
\begin{equation}
g_{00}^{-1} E^{2} + g_{11}^{-1} \Dot{r}^{2} + g_{33}L^{2} = 0, \label{eqsss}
\end{equation}
with $L$ is the angular momentum and $E$ being the energy. Next, Eq. (\ref{eqsss}) reads,
\begin{equation}
\Dot{r}^{2} = E^{2} - \left(  \frac{r^2 \left(1-\sqrt{\frac{4}{3} \alpha  \left(\Lambda +6 \mathcal{M} \left(\frac{1}{\left(l^2+r^2\right)^{3/2}}+\frac{\beta  e^{-\frac{r}{\lambda }} (\lambda +r)}{\lambda  r^3}\right)\right)+1}\right)}{2 \alpha }+1 \right)\left(  \frac{L^{2}}{r^{2}} \right),
\end{equation}
with $\overset{\nsim}{V} \equiv \left(  \frac{r^2 \left(1-\sqrt{\frac{4}{3} \alpha  \left(\Lambda +6 \mathcal{M} \left(\frac{1}{\left(l^2+r^2\right)^{3/2}}+\frac{\beta  e^{-\frac{r}{\lambda }} (\lambda +r)}{\lambda  r^3}\right)\right)+1}\right)}{2 \alpha }+1 \right)\left(  \frac{L^{2}}{r^{2}} \right)$, being the effective potential. To determine the critical radius, it is necessary to solve the equation $\partial \overset{\nsim}{V}/\partial r = 0$. In order to derive a viable solution, we adopt the assumption that $r$ is small. This consideration results in the emergence of four solutions for critical orbits. However, upon accounting for specific parameter values, only two of these solutions prove to be physically significant, exhibiting positive real values—namely, $r_{c_{\text{inner}}}$ and $r_{c_{\text{outer}}}$. For enhanced reader comprehension, we present Tables \ref{photonradius} and \ref{photonradius2}, which illustrate the behavior of these critical orbits.

Table \ref{photonradius} represents the behavior of $r_{c_{\text{inner}}}$. A discernible trend is observed: as the parameter $\mathcal{M}$ increases, there is a substantial decrease in the value of $r_{c_{\text{inner}}}$. Conversely, an opposing pattern emerges with an increase in the parameter $a$, leading to an increment in the values of $r_{c_{\text{inner}}}$. 

Furthermore, Table \ref{photonradius2} focuses on the analysis of $r_{c_{\text{outer}}}$. It is noteworthy that the behavior of $r_{c_{\text{outer}}}$ is fundamentally contrasting to that of $r_{c_{\text{inner}}}$. Specifically, as the parameter $\mathcal{M}$ increases, there is a simultaneous increase in the values of $r_{c_{\text{outer}}}$. Conversely, when the parameter $a$ experiences an increase, there is a corresponding decrease in the values of $r_{c_{\text{outer}}}$. Recently in the literature, there are many studies indicating more than one light ring within the context of modified gravity \cite{l1,l2}.

\begin{table}[!h]
\begin{center}
\begin{tabular}{c c  c ||| c c c } 
 \hline\hline
 $\mathcal{M}$ & $\alpha$ & $r_{c_{\text{inner}}}$ & $\mathcal{M}$ & $\alpha$ & $r_{c_{\text{inner}}}$  \\ [0.2ex] 
 \hline 
  0.0 & 0.10 & ------------ & 1.0 & 0.10 & 0.0890302  \\ 

  1.0 & 0.10 & 0.0890302 & 1.0 & 0.11 & 0.0979745  \\
 
  2.0 & 0.10 & 0.0444508 & 1.0 & 0.12 & 0.1069330  \\
 
  3.0 & 0.10 & 0.0296307 & 1.0 & 0.13 & 0.1159090  \\
 
  4.0 & 0.10 & 0.0222225 & 1.0 & 0.14 & 0.1249050  \\
 
  5.0 & 0.10 & 0.0177779 & 1.0 & 0.15 & 0.1339220  \\
 
  6.0 & 0.10 & 0.0148149 & 1.0 & 0.16 & 0.1429650  \\
 
  7.0 & 0.10 & 0.0126984 & 1.0 & 0.17 & 0.1520370   \\
 
 8.0 & 0.10 & 0.0111111 & 1.0 & 0.18 & 0.1611400   \\
 
 9.0 & 0.10 & 0.00987655 & 1.0 & 0.19 & 0.1702800    \\
 
 10.0 & 0.10 & 0.00888890 & 1.0 & 0.20 & 0.1794590    \\
 [0.2ex] 
 \hline \hline
\end{tabular}
\caption{\label{photonradius} The critical orbit values, denoted as $r_{c_{\text{inner}}}$, are presented across varying mass $M$ and parameter $\alpha$ values in the updated table.}
\end{center}
\end{table}

\begin{table}[!h]
\begin{center}
\begin{tabular}{c c  c ||| c c c } 
 \hline\hline
 $\mathcal{M}$ & $\alpha$ & $r_{c_{\text{outer}}}$ & $\mathcal{M}$ & $\alpha$ & $r_{c_{\text{outer}}}$  \\ [0.2ex] 
 \hline 
  0.0 & 0.10 & ---------- & 1.0 & 0.10 & 0.843597  \\ 

  1.0 & 0.10 & 0.843597 & 1.0 & 0.11 & 0.852279  \\
 
  2.0 & 0.10 & 0.989891 & 1.0 & 0.12 & 0.859277  \\
 
  3.0 & 0.10 & 1.055290 & 1.0 & 0.13 & 0.864865  \\
 
  4.0 & 0.10 & 1.093220 & 1.0 & 0.14 & 0.869258  \\
 
  5.0 & 0.10 & 1.118220 & 1.0 & 0.15 & 0.872625  \\
 
  6.0 & 0.10 & 1.136040 & 1.0 & 0.16 & 0.875104  \\
 
  7.0 & 0.10 & 1.149430 & 1.0 & 0.17 & 0.876803   \\
 
 8.0 & 0.10 & 1.159890 & 1.0 & 0.18 & 0.877815   \\
 
 9.0 & 0.10 & 1.168300 & 1.0 & 0.19 & 0.878212    \\
 
 10.0 & 0.10 & 1.175230 & 1.0 & 0.20 & 0.878057    \\
 [0.2ex] 
 \hline \hline
\end{tabular}
\caption{\label{photonradius2} The critical orbit values, denoted as $r_{c_{\text{outer}}}$, are presented across varying mass $M$ and parameter $\alpha$ values in the updated table.}
\end{center}
\end{table}

Within this section, our paramount aim is to grasp the intricacies of behavior resulting from the geodesic equation. This involves articulating it as follows:
\begin{equation}
\frac{\mathrm{d}^{2}x^{\mu}}{\mathrm{d}s^{2}} + \Gamma\indices{^\mu_\gamma_\sigma}\frac{\mathrm{d}x^{\gamma}}{\mathrm{d}s}\frac{\mathrm{d}x^{\sigma}}{\mathrm{d}s} = 0. 
\label{Gauss-Bonet}
\end{equation}
Considering Eq. \eqref{Gauss-Bonet}, the aforementioned equation gives rise to four partial differential equations

\begin{equation}
\frac{\mathrm{d}t^{\prime}}{\mathrm{d}s} = \frac{6 \alpha  r' t' \left(-\frac{r \left(1-\sqrt{\frac{4 \alpha  \Lambda }{3}+8 \alpha  M \left(\frac{1}{\left(l^2+r^2\right)^{3/2}}+\frac{\beta  e^{-\frac{r}{\lambda }} (\lambda +r)}{\lambda  r^3}\right)+1}\right)}{\alpha }-\frac{2 M \left(\frac{3 r^5}{\left(l^2+r^2\right)^{5/2}}+\frac{\beta  r^2 e^{-\frac{r}{\lambda }}}{\lambda ^2}+\frac{3 \beta  r e^{-\frac{r}{\lambda }}}{\lambda }+3 \beta  e^{-\frac{r}{\lambda }}\right)}{r^2 \sqrt{\frac{4 \alpha  \Lambda }{3}+8 \alpha  M \left(\frac{1}{\left(l^2+r^2\right)^{3/2}}+\frac{\beta  e^{-\frac{r}{\lambda }} (\lambda +r)}{\lambda  r^3}\right)+1}}\right)}{6 \alpha -3 r^2 \sqrt{\frac{4 \alpha  \Lambda }{3}+8 \alpha  M \left(\frac{1}{\left(l^2+r^2\right)^{3/2}}+\frac{\beta  e^{-\frac{r}{\lambda }} (\lambda +r)}{\lambda  r^3}\right)+1}+3 r^2},  \end{equation}

\begin{equation}
\begin{split}
\\
\frac{\mathrm{d}r^{\prime}}{\mathrm{d}s} =  & \frac{6 \alpha -3 r^2 \sqrt{\frac{4 \alpha  \Lambda }{3}+8 \alpha  M \left(\frac{1}{\left(l^2+r^2\right)^{3/2}}+\frac{\beta  e^{-\frac{r}{\lambda }} (\lambda +r)}{\lambda  r^3}\right)+1}+3 r^2}{12 \alpha  \left(\frac{r^2 \left(1-\sqrt{\frac{4 \alpha  \Lambda }{3}+8 \alpha  M \left(\frac{1}{\left(l^2+r^2\right)^{3/2}}+\frac{\beta  e^{-\frac{r}{\lambda }} (\lambda +r)}{\lambda  r^3}\right)+1}\right)}{2 \alpha }+1\right)^2} \times  \left\{  \right.\\
& \left.   \frac{2 M \left(\frac{3 r^5}{\left(l^2+r^2\right)^{5/2}}+\frac{\beta  r^2 e^{-\frac{r}{\lambda }}}{\lambda ^2}+\frac{3 \beta  r e^{-\frac{r}{\lambda }}}{\lambda }+3 \beta  e^{-\frac{r}{\lambda }}\right)}{r^2 \sqrt{\frac{4 \alpha  \Lambda }{3}+8 \alpha  M \left(\frac{1}{\left(l^2+r^2\right)^{3/2}}+\frac{\beta  e^{-\frac{r}{\lambda }} (\lambda +r)}{\lambda  r^3}\right)+1}} \right.\\
& \left. +\frac{r \left(1-\sqrt{\frac{4 \alpha  \Lambda }{3}+8 \alpha  M \left(\frac{1}{\left(l^2+r^2\right)^{3/2}}+\frac{\beta  e^{-\frac{r}{\lambda }} (\lambda +r)}{\lambda  r^3}\right)+1}\right)}{\alpha }
\right\} \left(r'\right)^2 \\
& - \frac{6 \alpha -3 r^2 \sqrt{\frac{4 \alpha  \Lambda }{3}+8 \alpha  M \left(\frac{1}{\left(l^2+r^2\right)^{3/2}}+\frac{\beta  e^{-\frac{r}{\lambda }} (\lambda +r)}{\lambda  r^3}\right)+1}+3 r^2}{12 \alpha} \times \left\{ \right. \\
& \left. \frac{2 M \left(\frac{3 r^5}{\left(l^2+r^2\right)^{5/2}}+\frac{\beta  r^2 e^{-\frac{r}{\lambda }}}{\lambda ^2}+\frac{3 \beta  r e^{-\frac{r}{\lambda }}}{\lambda }+3 \beta  e^{-\frac{r}{\lambda }}\right)}{r^2 \sqrt{\frac{4 \alpha  \Lambda }{3}+8 \alpha  M \left(\frac{1}{\left(l^2+r^2\right)^{3/2}}+\frac{\beta  e^{-\frac{r}{\lambda }} (\lambda +r)}{\lambda  r^3}\right)+1}} \right. \\
& \left. +\frac{r \left(1-\sqrt{\frac{4 \alpha  \Lambda }{3}+8 \alpha  M \left(\frac{1}{\left(l^2+r^2\right)^{3/2}}+\frac{\beta  e^{-\frac{r}{\lambda }} (\lambda +r)}{\lambda  r^3}\right)+1}\right)}{\alpha } \right\} \left(t'\right)^2
\\ & -\frac{r \left(-6 \alpha +3 r^2 \sqrt{\frac{4 \alpha  \Lambda }{3}+8 \alpha  M \left(\frac{1}{\left(l^2+r^2\right)^{3/2}}+\frac{\beta  e^{-\frac{r}{\lambda }} (\lambda +r)}{\lambda  r^3}\right)+1}-3 r^2\right)}{6 \alpha }\bigg((\theta ')^2+\sin ^2(\theta ) (\varphi ')^2\bigg),
\end{split}
\end{equation}

\begin{equation}
\frac{\mathrm{d} \theta^{\prime}}{\mathrm{d}s}  = - \sin (\theta ) \cos (\theta ) \left(\varphi '\right)^2-\frac{2 \theta ' r'}{r}, 
\end{equation}
and
\begin{equation}
 \frac{\mathrm{d} \phi^{\prime}}{\mathrm{d}s}    =    -\frac{2 \varphi ' \left(r'+r \theta ' \cot (\theta )\right)}{r}.
\end{equation}

\begin{figure}
    \centering
    \includegraphics[scale=0.45]{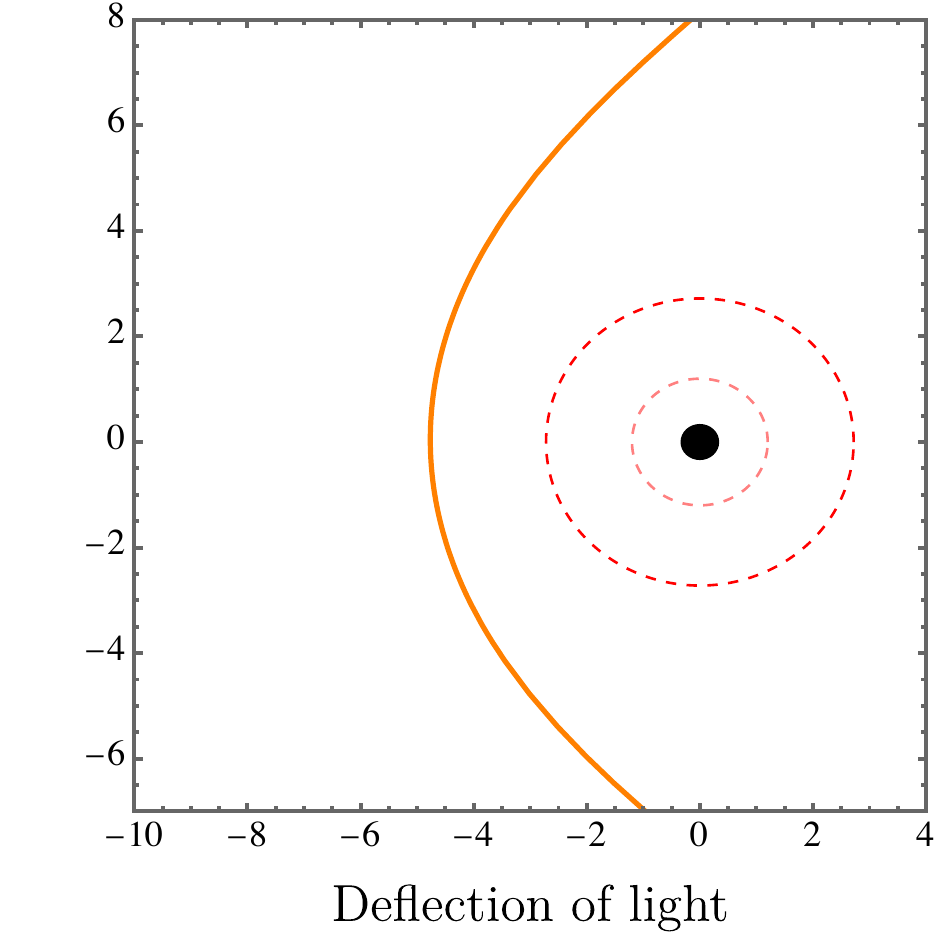}
    \caption{A particular representation of the deflection of light, considering $\alpha =0.1, \lambda =10^5, \Lambda = 10^{-5}, \mathcal{\beta}=0.1$, and $\mathcal{M} =1$. Here, the thick yellow line denotes the light; the dotted lines represent the photon sphere; and the red ones identify the shadows.}
    \label{light}
\end{figure}

In Fig. \ref{light}, it is depicted an illustrating deviation of light, incorporating parameters such as $\alpha = 0.1$, $\lambda = 10^5$, $\Lambda = 10^{-5}$, $\mathcal{\beta} = 0.1$, and $\mathcal{M} = 1$. In this representation, the thick yellow line signifies the path of light, the dashed yellow lines depict the photon sphere, and the dashed red lines indicate the regions of shadow.

\section{Null geodesics and Shadows}
This section takes care to study the null geodesic motion in $g_{\mu\nu}(x)$ background. The first step is to use the Lagrangian formalism \cite{chandrasekhar1998mathematical, Cruz:2004ts, Villanueva:2018kem} as the necessary concept to describe the geodesic process so that the Lagrangian in  the background $(\ref{solution})$ is given as:
\begin{equation}
   \mathcal{L}=-\frac{f(r)}{2}\,\dot{t}^2+\frac{f(r)^{-1}}{2}\,\dot{r}^2+\frac{r^2}{2}\,\mathcal{L}_{\Omega_{2}}
    \label{Lag}
\end{equation}
where $\mathcal{L}_{\Omega_{2}}$ is the angular Lagrangian, which can be given as such 
\begin{equation}
\mathcal{L}_{\Omega_{2}}=\dot{\theta}^2+\sin^2\theta\,\dot{\phi}^2.
    \label{E5-2}
\end{equation}
Here, the dot is a notation of the differentiation with respect to the affine parameter $\delta$ along the geodesic. Because the $\left(t,\phi\right)$ variables are cyclic, the underlying Lagrangian analysis $(\ref{Lag})$ is independent of them, and thus the associated conjugate momenta $\pi_q=\partial\mathcal{L}/\partial\dot{q}$ are conserved. Specifically, we have
\begin{align}
    \pi_t&=-f(r)\,\dot{t}\equiv-E,\\
    \pi_\phi&=r^2\,\sin^2\theta\,\dot{\phi}\equiv L,
    \label{E5-2}
\end{align}
where $E$ is a positive constant measuring the temporal invariance of the Lagrangian. Furthermore, this constant has no connection to energy due to the fact that the metric system $(\ref{solution})$ is not asymptotically flat, while $L$ is a constant, ensuring the conservation of the angular momentum.

A better approach for describing analytically the motion of a particle around the BH solution is carried out with the consideration of the Hamilton Jacobi formalism together with the Carter approach \cite{17}. In particular, the massless particle generates null geodesic motions (photon orbits). In this regard, the Hamilton-Jacobi equation is given in the following way:
\begin{eqnarray}
    \frac{\partial S}{\partial \delta}=-\frac{1}{2}g^{\mu\nu}\frac{\partial S}{\partial x^\mu}\frac{\partial S}{\partial x^\nu},
    \label{S}
\end{eqnarray}
where $S$ is the Jacobi action of the test particle and $\delta$ is the affine parameter. Incorporating the inverse metric component of the background $(\ref{solution})$ into the Eq.$(\ref{S})$, one can obtain
\begin{widetext}
    \begin{equation}
        -2\frac{\partial S}{\partial \delta}=-\frac{1}{f(r)}\left(\frac{\partial S_t}{\partial t}\right)^2+f(r)\left(\frac{\partial S_r}{\partial r}\right)^2+\frac{1}{r^2}\left(\frac{\partial S_\theta}{\partial \theta}+\frac{1}{\sin^2\theta}\left(\frac{\partial S_\phi}{\partial \phi}\right)^2\right).
        \label{-2}
    \end{equation}
\end{widetext}
It is worth noting that a separable solution for the Jacobi action is explicitly represented by
\begin{equation}
    S=\frac{1}{2}\mu ^2\delta-Et+L\phi+S_r(r)+S_\theta(\theta),
    \label{Jacobi}
\end{equation}
where $\mu$ is the rest mass of the test particle. In what follows, we restrict the study to dealing with mass particles such as photons $(\mu=0)$. As a result, implementing Jacobi's action $(\ref{Jacobi})$ on Eq. $(\ref{-2})$ leads to the following expression
\begin{widetext}
    \begin{align}
        0=\bigg\lbrace\frac{1}{f(r)}E^2-f(r)\left(\frac{\partial S_r}{\partial r}\right)^2-\frac{1}{r^2}\left(L^2+\mathcal{C}\right)\bigg\rbrace-\bigg\lbrace\frac{1}{r^2}\left(\frac{\partial S_\theta}{\partial\theta}\right)^2+\frac{1}{r^2}\left(L^2\cot^2\theta-\mathcal{C}\right)\bigg\rbrace,
    \end{align}
\end{widetext}
where $\mathcal{C}$ is the Carter constant. After executing some computations, the result provides the following independent couple of equations:
\begin{align}
    r^4f^2(r)\left(\frac{\partial S_r}{\partial r}\right)^2&=r^4\,E^2-r^2\left(L^2+\mathcal{C}\right)f(r),\\
    \left(\frac{\partial S_\theta}{\partial \theta}\right)^2&=\mathcal{C}-L^2\cot^2\theta.
\end{align}
In accordance with the Hamilton--Jacobi formalism, the appropriate set of the geodesic equation is given by
\begin{align}
\frac{\mathrm{d}t}{\mathrm{d}\delta}&=\frac{E}{f(r)},\label{1}\\
r^2\frac{\mathrm{d}r}{\mathrm{d}\delta}&=\pm\sqrt{\mathcal{R}},\label{2}\\
r^2\frac{\mathrm{d}\theta}{\mathrm{d}\delta}&=\pm\sqrt{\Theta},\label{3}\\
\frac{\mathrm{d}\phi}{\mathrm{d}\delta}&=\frac{L}{r^2\sin^2\theta}\label{4}
\end{align}
where the symbols $"+/-"$ signify the radial directions in which photons are moving, respectively, outward and inward. To complete the analysis view of the Hamilton-Jacobi formalism, the $\mathcal{R}$ and $\Theta$ are expressed as follows:
\begin{align}
    \mathcal{R}&=r^4\, E^2-r^2\left(L^2+\mathcal{C}\right)f(r)\\
    \Theta&=\mathcal{C}-L^2\cot^2\theta.
\end{align}
The photon's motion in the space-time $(\ref{solution})$ is controlled by Eqs. $(\ref{1})$-$(\ref{4})$.

To analyze in depth the shadow behaviors, it is required to define and carry an effective potential. This step certainly helps to show the shape of a black hole, which is entirely defined by the boundaries of its shadow and represents the apparent shape of the photon's unstable circular orbits. Thus, the effective potential is expressed in such a way as
\begin{equation}
    \left(\frac{\mathrm{d}r}{\mathrm{d}\delta}\right)^2+V_{eff}(r)=0.
    \label{E5-5}
\end{equation}
In other terms, the previous expression provides the equivalent of the following:
\begin{equation}
    V_{eff}(r)=f(r)\left\lbrace\frac{1}{r^2}\left(L^2+\mathcal{C}\right)-\frac{E^2}{f(r)}\right\rbrace.
    \label{E5-6}
\end{equation}

To specifically find the unstable circular orbit of the photons, which carry information about the boundary of the apparent shape of the black hole, the maximum of the effective potential provides the needed information through the following constraints \cite{Bardeen:1972fi}:
\begin{equation}
    V_{eff}=\frac{\mathrm{d}V_{eff}(r)}{\mathrm{d}r}\biggr\rvert_{r=r_{ph} }=0,\quad 
     \mathcal{R}=\frac{\mathrm{d}\mathcal{R}}{\mathrm{d}r}\biggr\rvert_{r=r_{ph} }=0.
    \label{cons}
\end{equation}
From which the photon sphere radius $r_{ph}$ is linked to the maximum effective potential for the spacetime black hole $(\ref{solution})$, which is the smallest value of the roots of the following equation:
\begin{eqnarray}\label{fp}
r_{ph}f'(r_{ph})-2f(r_{ph})=0
\label{E5-8}
\end{eqnarray}
where $f'(r)$ is a notation of the differentiation $\frac{\partial f}{\partial r}$ and in terms of the parameter space of the BH system. Solving Eq. (\ref{fp}) to generate the explicit expression for the photon sphere seems analytically difficult, which leads to the application of the numerical approach. As a consequence, Tab. \ref{rp} collects information on the photon sphere $r_{ph}$ with respect to several fixed parameters.
\begin{table*}[ht!]
    \begin{tabular}{lcccc} 
    \hline\hline
        $r_{ph} $  \,& $\lambda$ \,& $\Lambda$ \,&  $\alpha$ \,& $\beta$ \
       \\
       \hline
       1.16674\,&\,0.1 \,&\, 0.001 \,&\, 0.85 \,&\, 0.1\\
       2.01037 \,&\, 0.1 \,&\, 0.002 \,&\, 0.85 \,&\,0.1 \\
       2.10125 \,&\, 0.1 \,&\, 0.002 \,&\, 0.5  \,&\, 0.2\\
       1.99236 \,&\, 0.7 \,&\, 0.002 \,&\, 0.85  \,&\, 0.2 \\
           \hline\hline
    \end{tabular}
     \caption{Numerical sets of the photon sphere $r_{ph} $ with $M =1$ and $\ell_0=0.2$.}
      \label{rp}
    \end{table*}

The present step consists of defining the shape and size of the black hole while considering space-time $(\ref{solution})$. For this, the impact parameters $\xi$ and $\eta$ are needed and related to the constants of motion $E$, $L$, and $\mathcal{C}$ by means of the following expressions:
\begin{eqnarray}
    \xi=\frac{L}{E},\quad \eta=\frac{\mathcal{C}}{E^2},
\end{eqnarray}
where the effective potential and the radial function become expressed in terms of these impact parameters, as shown in the following
\begin{align}
    V_{eff}&=E^2\bigg\lbrace\frac{f(r)}{r^2}\left(\xi+\eta\right)-1\bigg\rbrace,\label{cons1}\\
    \mathcal{R}&=E^2\bigg\lbrace r^4-r^2\left(\xi+\eta\right)f(r)\bigg\rbrace.\label{cons2}
\end{align}
Afterward, injecting Eqs. $(\ref{cons1})$-$(\ref{cons2})$ into Eq. $(\ref{cons})$ leads to obtaining an equation for two unknowns, $\xi$ and $\eta$, in the following compact form:
\begin{equation}
\eta+\xi^2=\frac{4r^3_{ph}}{2r_{ph}f(r_{ph})+r^2_{ph}f'(r_{ph})}.
\end{equation}
As a result, it may be noted that the photon sphere $r_{ph}$ has the dimensions of the length, while the quantity $\eta+\xi^2$ has the dimensions of the length square, describing a two-dimensional shadow geometry. 
\begin{figure*}[tbh!]
      	\centering{
       \includegraphics[scale=0.59]{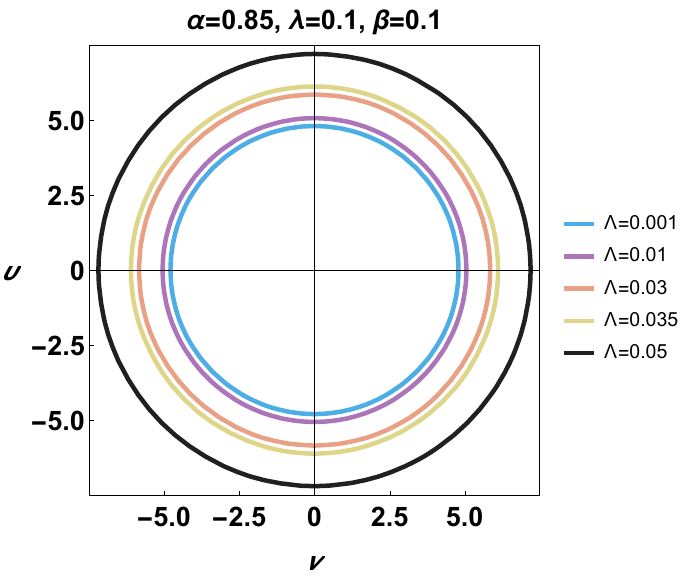} \hspace{2mm}
      	\includegraphics[scale=0.57]{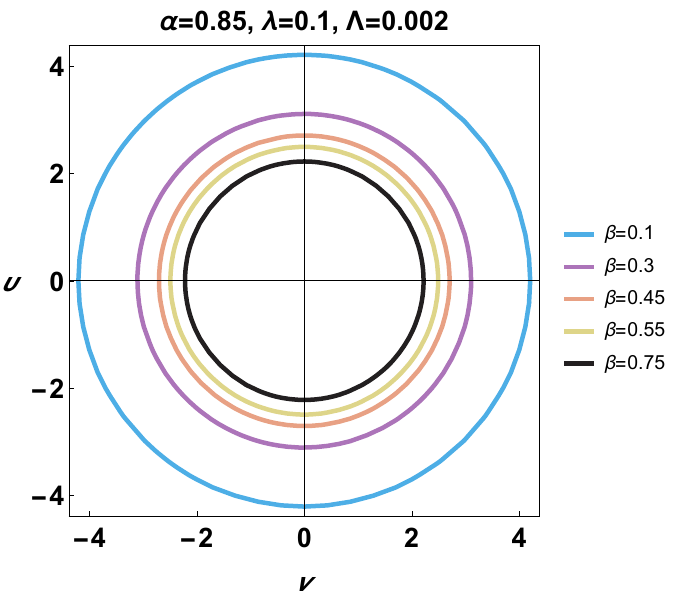} \hspace{2mm}
      }
       \centering{ 
       \includegraphics[scale=0.57]{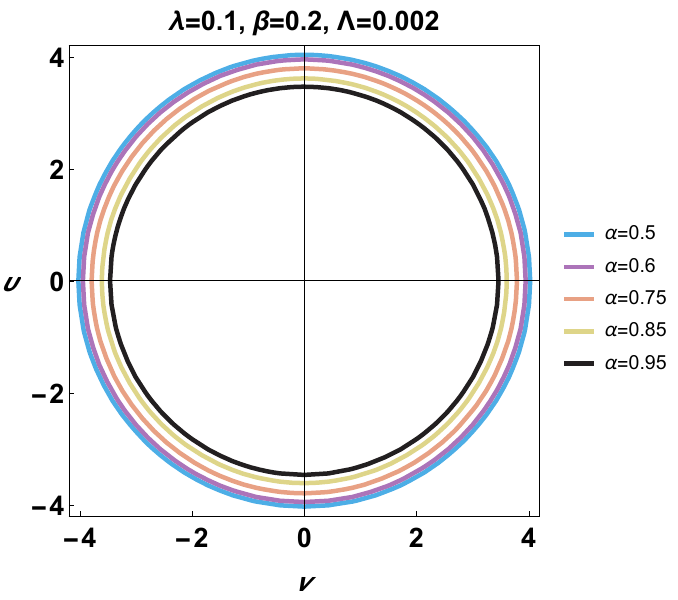}\hspace{1.5mm}
      \includegraphics[scale=0.57]{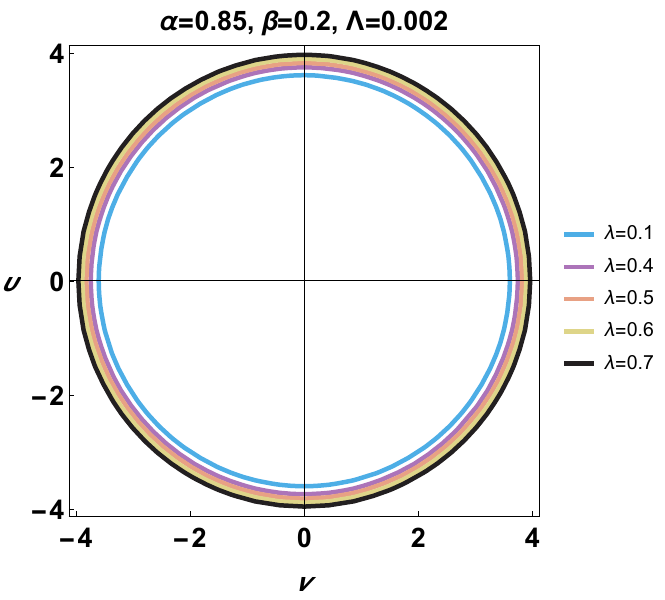} \hspace{1mm}
       }
      	\caption{The geometrical shape of the shadow and the shadow radius of the BH solution in the celestial plane for several values of $\Lambda$, $\beta$, $\alpha$ and $\lambda$ with $M= 1$ and $\ell_0= 0.2$}
      	\label{fig4}
      \end{figure*}
      
 Now, we turn to reveal the visualization of the black hole shadow, i.e., the geometrical quantity on a celestial plane along the coordinates $\nu$ and $\upsilon$. Therefore, the celestial coordinates are given according to \cite{Vazquez:2003zm} by
\begin{align}
\nu&=\underset{r_O\rightarrow\infty}{lim}\left(-r_O^2\sin\theta_O\frac{\mathrm{d}\phi}{\mathrm{d}r}\biggr\rvert_{(r_O,\theta_O)}\right)\\
\upsilon&=\underset{r_O\rightarrow\infty}{lim}\left(r_O^2\frac{\mathrm{d}\theta}{\mathrm{d}r}\biggr\rvert_{(r_O,\theta_O)}\right),
\label{E5-15}
\end{align}
where $r_O$ denotes the distance between the black hole and the observer. To be more concrete, we look at null geodesic motion in the equatorial plane $\theta=\pi/2$ leading to $\nu=-\xi$ and $\upsilon=\pm\sqrt{\eta}$. As a consequence, this outcome presents a two-dimensional geometry governed by the shadow radius expressed in the following way:
\begin{equation}
    R_s^2\equiv \eta+\xi^2=\nu^2+\upsilon^2
\end{equation}
which is nothing more than the shadow radius $R_s$ in celestial coordinates. Bearing in mind that the shadow shape for non-rotating (static) black holes is a circle with a radius of $R_s$.

The practical step involved in analyzing the appropriate shadow behavior of the BH solution is perfectly depicted in Fig. \ref{fig4}. So, proceeding with the shadow behavior is mainly carried out by analyzing the geometrical behavior of the shadow radius. As it is observed, the size of the BH shadow is proportional to the variation of the parameters $\lambda$ and $\Lambda$, while disproportional to $\alpha$ and $\beta$.

\section{Quasinormal modes}

In the ringdown phase, the extraordinary phenomenon of \textit{quasinormal} modes unfolds, revealing oscillation patterns that remain remarkably steadfast against initial perturbations. These modes intricately capture the inherent essence of the system, stemming from the natural oscillations of spacetime and transcending the influence of specific initial conditions. In contrast to the more confined nature of \textit{normal} modes within closed systems, \textit{quasinormal} modes characterize open systems, gradually releasing energy through the emission of gravitational waves. Mathematically, these modes are eloquently expressed as poles of the complex Green function.

Unraveling solutions to the wave equation within a system dictated by the background metric \(g_{\mu\nu}\) is crucial for determining their frequencies. Nevertheless, the quest for analytical solutions to these modes poses formidable challenges. The scientific community has explored diverse methodologies to surmount this complexity, with the WKB (Wentzel--Kramers--Brillouin) method emerging as a prominent approach. Its roots trace back to the pioneering work of Will and Iyer \cite{Iyer:1986np,Iyer:1986nq}. Subsequent refinements have elevated the method, with Konoplya extending it to the sixth order \cite{Konoplya:2003ii}, and Matyjasek and Opala pushing the boundaries to the thirteenth order \cite{Matyjasek:2017psv}.

\subsection{Scalar perturbations}

In our computations, we focus on scrutinizing perturbations utilizing the Klein-Gordon equation within the context of curved spacetime, specifically employing the scalar field
\begin{equation}
\frac{1}{\sqrt{-g}}\partial_{\mu}(g^{\mu\nu}\sqrt{-g}\partial_{\nu}\Phi) = 0.\label{Klein}
\end{equation}
While exploring the intriguing realm of \textit{backreaction} effects is tempting in this context, this manuscript deliberately shifts its focus to other aspects. Our primary concentration lies in the meticulous examination of the scalar field as a minor perturbation. The inherent spherical symmetry in the scenario allows us to decompose the scalar field in a specific manner, as elucidated in greater detail below
\begin{equation}
\Phi(t,r,\theta,\varphi) = \sum^{\infty}_{l=0}\sum^{l}_{m=-l}r^{-1}\Psi_{lm}(t,r)Y_{lm}(\theta,\varphi).\label{posss}
\end{equation}
In this context, the spherical harmonics are denoted by $Y_{lm}(\theta,\varphi)$. Upon integrating the scalar field decomposition, as presented in Eq. \eqref{posss}, into Eq. \eqref{Klein}, the equation assumes a Schrödinger--like form. This transformation bestows upon the equation wave--like attributes, uniquely aligning it with the nuances essential for our analytical investigation
\begin{equation}
-\frac{\partial^{2} \Psi}{\partial t^{2}}+\frac{\partial^{2} \Psi}{\partial r^{*2}} + V_{eff}(r^{*})\Psi = 0.\label{schorger}
\end{equation}

The potential \(V_{eff}\), commonly known as the \textit{Regge--Wheeler} potential or the effective potential, encapsulates crucial details about the geometry of the black hole. To enhance our analysis, we introduce the tortoise coordinate \(r^{*}\), spanning the entire spacetime and approaching \(\pm \infty\) as \(r^{*}\) extends. This relationship is represented by \(\mathrm{d} r^{*} = \sqrt{[1/f(r)^{2}]}\mathrm{d}r\). Through algebraic rearrangements, we can concisely express the effective potential as:
\begin{equation}
\begin{split}
V_{eff} = & f(r)\left[ -\frac{2 M r^2 \left(-\frac{3 r}{\left(l^2+r^2\right)^{5/2}}-\frac{3 \beta  e^{-\frac{r}{\lambda }} (\lambda +r)}{\lambda  r^4}-\frac{\beta  e^{-\frac{r}{\lambda }} (\lambda +r)}{\lambda ^2 r^3}+\frac{\beta  e^{-\frac{r}{\lambda }}}{\lambda  r^3}\right)}{r\sqrt{\frac{4}{3} \alpha  \left(\Lambda +6 M \left(\frac{1}{\left(l^2+r^2\right)^{3/2}}+\frac{\beta  e^{-\frac{r}{\lambda }} (\lambda +r)}{\lambda  r^3}\right)\right)+1}} \right.\\
& \left. + \frac{r \left(1-\sqrt{\frac{4}{3} \alpha  \left(\Lambda +6 M \left(\frac{1}{\left(l^2+r^2\right)^{3/2}}+\frac{\beta  e^{-\frac{r}{\lambda }} (\lambda +r)}{\lambda  r^3}\right)\right)+1}\right)}{\alpha \,r} + \frac{l (l+1)}{r^2} \right].\label{potteential}
\end{split}
\end{equation}

Notably, a barrier--like structure manifests prominently when the cosmological constant $\Lambda$, coupling constant $\lambda$, and scalar field parameter $\alpha$ all take positive values. It is worth observing that as the angular momentum quantum number $l$ increases, there is a corresponding elevation in the height of $V_{eff}$.

To compute the \textit{quasinormal} modes, we concentrate on the WKB method. Our primary aim is to derive stationary solutions for the system. To achieve this, we propose the representation \(\Psi(t,r) = e^{-i\omega t} \psi(r)\), where \(\omega\) signifies the frequency. This choice enables us to seamlessly isolate the time--independent aspect of Eq. \eqref{schorger}, as delineated below:
\begin{equation}
\frac{\partial^{2} \psi}{\partial r^{*2}} - \left[  \omega^{2} - V_{eff}(r^{*})\right]\psi = 0.\label{ftimeindependent}
\end{equation}

To effectively address Eq. \eqref{ftimeindependent}, it is imperative to meticulously account for the relevant boundary conditions. In our specific scenario, solutions adhering to these conditions are discerned by their distinctively exclusive purely ingoing behavior in the vicinity of the horizon
\[
    \psi^{\text{in}}(r^{*}) \sim 
\begin{cases}
    C_{l}(\omega) e^{-i\omega r^{*}} & ( r^{*}\rightarrow - \infty)\\
    A^{(-)}_{l}(\omega) e^{-i\omega r^{*}} + A^{(+)}_{l}(\omega) e^{+i\omega r^{*}} & (r^{*}\rightarrow + \infty).\label{boundarycon}
\end{cases}
\]

The intricate complex constants \(A^{(+)}_{l}(\omega)\), \(C_{l}(\omega)\), and \(A^{(-)}_{l}(\omega)\) constitute the fundamental components of our subsequent analysis. They play a pivotal role in the examination of the \textit{quasinormal} modes of a black hole, characterized by frequencies \(\omega_{nl}\) that satisfy the condition \(A^{(-)}_{l}(\omega_{nl})=0\). These modes manifest a unique behavior, presenting as purely outgoing waves at spatial infinity and exclusively ingoing waves near the event horizon. The integers \(n\) and \(l\) delineate the overtone and multipole numbers, respectively. Furthermore, it is noteworthy that the spectrum of \textit{quasinormal} modes is grounded in the eigenvalues of Eq. \eqref{ftimeindependent}. To investigate these frequencies formally, we employ the WKB method, a semi-analytical approach reminiscent of quantum mechanics.

Additionally, the WKB approximation, originally introduced by Schutz and Will \cite{1985ApJ...291L..33S}, has become an indispensable method for determining \textit{quasinormal} modes, particularly in the context of studying particle scattering around black holes. Over the years, this technique has undergone refinement, marked by substantial contributions from Konoplya \cite{Konoplya:2003ii,Konoplya:2004ip}. It is essential to recognize, however, that the applicability of this method is contingent on the potential assuming a barrier-like form and leveling off to constant values as \(r^{*} \to \pm \infty\). By aligning the solution power series with the peak potential turning points, the \textit{quasinormal} modes can be accurately derived \cite{Santos:2015gja,Santos:2015gja}. Given these premises, the sixth--order WKB formula is expressed as follows:
\begin{equation}
\frac{i(\omega^{2}_{n}-V_{0})}{\sqrt{-2 V^{''}_{0}}} - \sum^{6}_{j=2} \Lambda_{j} = n + \frac{1}{2}.
\end{equation}

In essence, Konoplya's formulation for the \textit{quasinormal} modes comprises various essential components. Specifically, the term \(V^{''}_{0}\) denotes the second derivative of the potential, calculated at its zenith \(r_{0}\). Additionally, the constants \(\Lambda_{j}\) are influenced by the effective potential and its derivatives at this peak. Noteworthy advancements in this field have recently unveiled a 13th--order WKB approximation, pioneered by Matyjasek and Opala \cite{Matyjasek:2017psv}, markedly enhancing the precision of \textit{quasinormal} frequency computations.

Take note that the \textit{quasinormal} frequencies associated with the scalar field possess a negative imaginary component. This characteristic indicates that these modes undergo exponential decay over time, representing the dissipation of energy through scalar waves. This observation is consistent with previous investigations into perturbations in spherically symmetric configurations, encompassing scalar, electromagnetic, and gravitational perturbations \cite{Konoplya:2011qq,Berti:2009kk,Heidari:2023bww,2023InJPh.tmp..228C,q1,q2,q3}. In Tabs. \ref{scalar1}, \ref{scalar2} and \ref{scalar3}, we represent the values of the quasinormal modes for the scalar perturbations for different values of $\alpha$. Under certain configuration of the system, unstable modes appear.

\begin{table}[!h]
\begin{center}
\caption{\label{scalar1} Employing the sixth--order WKB approximation, we examine the quasinormal frequencies associated with scalar perturbations across different values of \(\alpha\), with a specific focus on cases where \(l=0\).}
\begin{tabular}{c| c | c | c} 
 \hline\hline\hline 
  $\alpha$    & $\omega_{0}$ & $\omega_{1}$ & $\omega_{2}$  \\ [0.2ex] 
 \hline 
 0.10  & 1.32496 - 1.03427$i$ & 20.0727 - 1.14659$i$ &  94.801 - 4.70918$i$  \\
 
 0.11  & 0.86332 - 1.8318$i$ & 12.1996 - 1.26940$i$  &  67.9051 - 2.94847$i$  \\
 
 0.12  &  0.877716 - 1.55745$i$  & 13.833 - 1.06241$i$  &  71.5674 - 3.10621$i$  \\
 
 0.13 & 0.9181 - 1.29472$i$ & 14.9734 - 0.973842$i$ & 74.189 - 3.33687$i$  \\
 
 0.14 & 0.989237 - 1.04963$i$ & 15.8141 - 0.941273$i$ & 76.1426 - 3.59839$i$   \\

  0.15 & 1.09109 - 0.834005$i$ & 16.4585 - 0.93852$i$ & 77.6471 - 3.86977$i$   \\
   [0.2ex] 
 \hline \hline \hline 
\end{tabular}
\end{center}
\end{table}

\begin{table}[!h]
\begin{center}
\caption{\label{scalar2} Employing the sixth--order WKB approximation, we examine the quasinormal frequencies associated with scalar perturbations across different values of \(\alpha\), with a specific focus on cases where \(l=1\).}
\begin{tabular}{c| c | c | c} 
 \hline\hline\hline 
  $\alpha$    & $\omega_{0}$ & $\omega_{1}$ & $\omega_{2}$  \\ [0.2ex] 
 \hline 
 0.10  & 37.9636 - 1.22204$\times$ $10^7$$i$ & 116.088 - 4.4149$\times$$10^7$$i$ & 252.091 - 1.13043$\times$$10^8$ \\
 
 0.11  & \text{Unstable}  & \text{Unstable}   & \text{Unstable}   \\
 
 0.12  &  3.80996$\times 10^6$ - 23.1870$i$  & 1.3713$\times 10^7$ - 70.7676$i$  &  3.50226$\times 10^7$ - 153.528$i$ \\
 
 0.13 & \text{Unstable} & \text{Unstable} & \text{Unstable} \\
 
 0.14 & 1.1673$\times 10^6$ - 16.2341$i$  & 4.21718$\times 10^6$ - 49.6392$i$ & 1.07982$\times 10^7$ - 107.7900$i$  \\

 0.15 & 2.14988$\times 10^6$ - 13.3484$i$  & 8.24887$\times 10^5$ - 5.93913$i$  & 2.14988$\times 10^6$ - 13.3484$i$  \\
   [0.2ex] 
 \hline \hline \hline 
\end{tabular}
\end{center}
\end{table}

\begin{table}[!h]
\begin{center}
\caption{\label{scalar3} Employing the sixth-order WKB approximation, we examine the quasinormal frequencies associated with scalar perturbations across different values of \(\alpha\), with a specific focus on cases where \(l=2\).}
\begin{tabular}{c| c | c | c} 
 \hline\hline\hline 
  $\alpha$    & $\omega_{0}$ & $\omega_{1}$ & $\omega_{2}$  \\ [0.2ex] 
 \hline 
 0.10  & 2.30133 - 4.53642$\times 10^{4} i$ & 90.5567 - 1.50169$\times 10^7$$i$ & 196.156 - 3.84392$\times 10^7$ $i$ \\
 
 0.11  & \text{Unstable} & \text{Unstable}   &  \text{Unstable}  \\
 
 0.12  &  762841 - 6.00149$i$  & 2.72701$10^6$ - 18.4951$i$  &  6.93186$\times 10^6$ - 40.3868$i$ \\
 
 0.13 & 8.24779 - 8.76037$\times 10^5$ $i$ & 25.1808 - 3.15653$\times 10^6$$i$ & 54.6363 - 8.06777$\times10^6$$i$ \\
 
 0.14 & 137356 - 5.25372$i$  & 510230. - 15.8237$i$ & 1.3302$\times 10^6$ - 34.0112$i$  \\

  0.15 & 2.30133 - 4.53642$\times 10^4$$i$ & 6.96983 - 171094$i$ & 15.0289 - 450159$i$  \\
   [0.2ex] 
 \hline \hline \hline 
\end{tabular}
\end{center}
\end{table}


\subsection{Vectorial perturbations}

In this section, we advance our exploration into the propagation of the electromagnetic field. To accomplish this, we revisit the wave equations governing a test electromagnetic field
\begin{equation}
\frac{1}{\sqrt{-g}}\partial_{\nu}\left[ \sqrt{-g} g^{\alpha \mu}g^{\sigma \nu} \left(A_{\sigma,\alpha} -A_{\alpha,\sigma}\right) \right]=0.
\end{equation}
The four--potential, represented as \(A_{\mu}\), can be expressed through an expansion in 4-dimensional vector spherical harmonics as follows:
\begin{small}
\begin{align}\notag
& A_{\mu }\left( t,r,\theta ,\phi \right)  \nonumber \\
&=\sum_{\ell ,m} 
\begin{bmatrix} 
f(t,r)Y_{\ell m}\left( \theta ,\phi \right) \\
h(t,r)Y_{\ell m}\left( \theta ,\phi \right) \\
\frac{a(t,r)}{\sin \left( \theta \right) }\partial _{\phi }Y_{\ell
m}\left( \theta ,\phi \right) + k(t,r)\partial _{\theta }Y_{\ell m}\left( \theta ,\phi \right)\\
-a\left( t,r\right) \sin \left( \theta \right) \partial _{\theta }Y_{\ell
m}\left( \theta ,\phi \right)+ k(t,r)\partial _{\varphi }Y_{\ell m}\left( \theta ,\phi \right)
\end{bmatrix}.%
\end{align}%
\end{small}

In the context of this expansion, \(Y_{\ell m}(\theta,\phi)\) represents the spherical harmonics. It is pertinent to note that the first term on the right--hand side exhibits a parity of \((-1)^{\ell +1}\) (referred to as the axial sector), while the second term holds a parity of \((-1)^\ell\) (known as the polar sector). Upon direct substitution of this expansion into the Maxwell equations, a second--order differential equation governing the radial component can be rigorously derived \cite{Toshmatov:2017bpx}
\begin{equation}
\frac{\mathrm{d}^{2}\Psi \left( r_{\ast }\right) }{\mathrm{d}r_{\ast }^{2}}+\left[ \omega
^{2}-V_{E}\left( r_{\ast }\right) \right] \Psi \left( r_{\ast }\right) =0.
\end{equation}%

In both the axial and polar sectors, we derive a second-order differential equation governing the radial component, where the tortoise coordinate is defined as \(r_{\ast} = \int f^{-1}(r)dr\). The mode \(\Psi(r_{\ast})\) represents a linear combination of the functions \(a(t,r)\), \(f(t,r)\), \(h(t,r)\), and \(k(t,r)\). However, the specific functional dependence varies depending on the parity. In the case of the axial sector, the mode is expressed as follows:
\begin{equation}
a(t,r)=\Psi \left( r_{\ast }\right).
\end{equation}
In contrast, for the polar sector, it is expressed as:
\begin{equation}
\Psi \left( r_{\ast }\right) =\frac{r^{2}}{%
\ell (\ell +1)}\left[ \partial _{t}h(t,r)-\partial _{r}f(t,r)\right].
\end{equation}
The associated effective potential in our case is determined as:
\begin{eqnarray}
    V_{eff}(r)=f(r) \left(\frac{l(l+1)}{r^2}\right).
\end{eqnarray}

In Tabs. \ref{vectorial2} and \ref{vectorial3}, we represent the values of the quasinormal modes of the vectorial perturbations for different values of $\alpha$. Here, we notice that for some values of $\alpha$ unstable modes give rise to.

\begin{table}[!h]
\begin{center}
\caption{\label{vectorial2} Employing the sixth--order WKB approximation, we examine the quasinormal frequencies associated with vectorial perturbations across different values of \(\alpha\), with a specific focus on cases where \(l=1\).}
\begin{tabular}{c| c | c | c} 
 \hline\hline\hline 
  $\alpha$    & $\omega_{0}$ & $\omega_{1}$ & $\omega_{2}$  \\ [0.2ex] 
 \hline 
 0.10  & 212.252 - 0.000331183$i$ &  763.062 - 0.000806585$i$    & 1947.29 - 0.00164103$i$  \\ 
 
 0.11  & 0.000543762 - 78.7522$i$ & 0.00101029 - 277.368$i$ & 0.00175005 - 697.531$i$ \\
 
 0.12  & \text{Unstable} &  \text{Unstable}  &  \text{Unstable}  \\
 
 0.13  &  60.1637 - 0.00225955$i$ & 216.598 - 0.00611754$i$ &  553.286 - 0.0128075$i$ \\
 
 0.14 & 59.6741 - 0.0031472$i$  & 215.189 - 0.00885143$i$ & 550.307 - 0.0187747$i$ \\
 
 0.15 & 0.00455997 - 11.70844$i$ & 0.0106379 - 39.3753$i$ & 0.0216639 - 95.3931$i$  \\
   [0.2ex] 
 \hline \hline \hline 
\end{tabular}
\end{center}
\end{table}

\begin{table}[!h]
\begin{center}
\caption{ \label{vectorial3} Employing the sixth--order WKB approximation, we examine the quasinormal frequencies associated with vectorial perturbations across different values of \(\alpha\), with a specific focus on cases where \(l=2\).}
\begin{tabular}{c| c | c | c} 
 \hline\hline\hline 
  $\alpha$    & $\omega_{0}$ & $\omega_{1}$ & $\omega_{2}$  \\ [0.2ex] 
 \hline 
 0.10  & \text{Unstable} &  \text{Unstable}  & \text{Unstable}  \\ 
 
 0.11  & \text{Unstable} & \text{Unstable} & \text{Unstable} \\
 
 0.12  & 8.89064 - 0.0112879$i$ & 33.3585 - 0.0246925$i$  &  87.5577 - 0.0477644$i$  \\
 
 0.13  &  19.644 - 0.00288482$i$ & 70.8999 - 0.00468075$i$ &  181.445 - 0.00780043$i$ \\
 
 0.14 & 12.8029 - 0.0054649$i$ & 45.8716 - 0.0105014$i$ & 116.827 - 0.0193291$i$ \\
 
 0.15 & 6.32386 - 0.00447958$i$ & 22.2207 - 0.00107095$i$ & \text{Unstable}  \\
   [0.2ex] 
 \hline \hline \hline 
\end{tabular}
\end{center}
\end{table}

\section{Conclusion}

Dealt with Yukawa--modified potential as a contribution of matter, in the context of EGB gravity, it had paved the way to inspect modeled black hole solutions. Some interesting geometrical constraints were practically encoded by the parameters $\ell_0$, $\beta$, $\lambda$, and the GB coupling constant $\alpha$. In particular, the spacetime geometry was modified at a small distance by the change of the parameter $\ell_0$, while at large distances, the spacetime was modified with the given parameter set $(\beta, \lambda, \alpha$), with the reason that $\beta$ modified Newton’s law of gravity and could mimic the dark matter effect, as well as the GB coupling constant, which guaranteed the validity of the EGB theory. The recent study gained fruitful investigations such as local thermal stability, critical orbits, shadow behaviors, and QNMs. Thus, the black hole solution was locally thermally stable. Moreover, the shadow behavior showed some dependencies between the space parameter of the black hole system and the size of the black hole shadow.

\section*{Acknowledgments}
\hspace{0.5cm}

A. A. Araújo Filho would like to thank Fundação de Apoio à Pesquisa do Estado da Paraíba (FAPESQ) and Conselho Nacional de Desenvolvimento Cientíıfico e Tecnológico (CNPq)  -- [150891/2023-7] for the financial support. Most of the calculations were performed by using the \textit{Mathematica} software. This work was partly supported by the Ministry of Science and Higher Education of the Republic of Kazakhstan, Grant AP14870191.


\bibliographystyle{ieeetr}
\bibliography{main}
\end{document}